%% file: main.tex
\documentclass[conference]{IEEEtran}

\usepackage[T1]{fontenc}
\usepackage[utf8]{inputenc}
\usepackage{graphicx}
\usepackage{booktabs}
\usepackage{array}
\usepackage{amsmath}
\usepackage{xcolor}
\usepackage{tikz}
\usetikzlibrary{arrows.meta,positioning,fit,calc,shapes.geometric}
\usepackage[hidelinks]{hyperref}
\usepackage[capitalize,noabbrev]{cleveref}
\usepackage{multirow}
\usepackage{listings}

\newcommand{\sys}{\textsc{moefs}\xspace}
\usepackage{xspace}
\newcommand{\tblhead}[1]{\textbf{#1}}
\newcommand{\yes}{\ding{51}}
\newcommand{\no}{\ding{55}}
\usepackage{pifont}

\begin{document}

\title{Searching for Plans You Can Actually Build:\\
A Realizability-Aware Full-Space Optimizer for MoE Training and Serving}
\author{\IEEEauthorblockN{Quan Yuan and Jie Zhao}%
\IEEEauthorblockA{Hunan University \\ \{quanyy, jiezhao\}@hnu.edu.cn}}

\maketitle

\begin{abstract}
\input{sections/00-abstract}

\end{abstract}

\input{sections/01-intro}

\input{sections/02-background}

\input{sections/03-design}
\input{sections/04-implementation}

\input{sections/05-evaluation}

\input{sections/06-discussion}

\input{sections/07-related}

\input{sections/08-conclusion}

\bibliographystyle{IEEEtran}
\bibliography{refs}

\appendix
\input{sections/09-appendix}

\end{document}

%% file: sections/00-abstract.tex
Mixture-of-Experts (MoE) systems split a program's plan space in two: the
space a cost model can rank, and the smaller space a real toolchain can
actually build. Automatic optimizers rank the first and silently assume the
two coincide---so they can return a plan that is optimal on paper and
impossible to emit. We present \sys, a realizability-aware full-space
optimizer for MoE training and serving that makes deployment realizability a
first-class search constraint. \sys closes a three-tier search over
parallelism, schedule, and kernels; it emits both a Megatron training stack
and an SGLang serving stack from a single plan; and it prices, rather than
forbids, the realization overheads it measures. We evaluate across two
hardware generations. On 2$\times$RTX4090, the searched training plan edges the
strongest hand-tuned baseline by +0.9\% (and clears the 0.98$\times$ acceptance
bar by +2.9\%);
on 8$\times$H800, the searched serving plan matches the hand-tuned
configuration at a 1.0304 throughput ratio.
We hold failures to the same standard: on 8$\times$H800 training, the searched
plan is a computable, honest FAIL at 0.9338 of the best hand-tuned throughput,
losing on a single schedule flag.
All predictions are pre-registered in a frozen, artifact-hashed adjudication
file before the H800 runs, and every outcome is reported as-is.

%% file: sections/01-intro.tex
\section{Introduction}
\label{sec:intro}

Ask a state-of-the-art automatic MoE optimizer for the best training plan on
eight H800 GPUs, and it may hand you a plan you cannot build. This is not
hypothetical. Run our own optimizer on exactly that question and, before any
measurement, its top-ranked training plan was \texttt{tp1/pp1/ep8} with a
DualPipe schedule, at a predicted 24{,}712.9 tokens/s/GPU---a plan that no
available toolchain can emit.
Megatron core v0.18.0 has no DualPipe schedule, so code generation halts with
a \texttt{NotImplementedError} before a single training step runs.
The number-one plan, by the model's own ranking, does not exist as runnable
code.

This is not an accuracy bug to be tuned away: after calibration the same plan
ranks even higher, at 27{,}758.2 tokens/s/GPU.
It is a category error in how automatic MoE optimizers are built. Such systems
search the space a cost model can rank and implicitly assume that this space
is the space the toolchain can actually build. The two spaces are different,
and the gap between them is systematically ignored. A plan can be optimal
under every analytical model---memory, compute, and communication---and still
have no realization on the target stack, because integer-divisibility rules,
unmapped schedule classes, sharding semantics, and missing kernels quietly
remove it from the buildable set. For an engineer this is not a slightly
wrong number but a command that throws before the first step, or a job that
runs out of memory once launched---after the optimizer has already declared it
best. The serving side tells the same story: the
pre-registered top serving plan (\texttt{ep4}) is unrealizable as modeled, and
the plan that actually ran fastest (\texttt{tp8}) was not even in the
pre-registered top five.

We present \sys, a realizability-aware full-space optimizer for MoE training
and serving~\cite{deepseek2024v3,kimi2025k2}. \sys treats deployment
realizability as a first-class search constraint rather than a post-hoc
filter: a realizability predicate reuses the code emitter itself as a dry
run---one source of truth, not a parallel rule base---and rejects any plan the
emitter cannot lower to runnable code. Around this predicate \sys closes a
loop. It runs a three-tier search over parallelism, schedule, and kernels;
scores candidates with dual-fidelity cost models (analytic and simulator);
emits both a Megatron training stack~\cite{shoeybi_megatronlm} and an SGLang
serving stack~\cite{sglang2025largeep} from one plan, unifying training and
serving; and prices, rather than forbids, the realization overheads it
measures on the target hardware. Pricing rather than forbidding keeps a plan
in play once its realized cost has been honestly accounted for, instead of
banning it on a heuristic. Folded into the search, the predicate turns
the H800 training frontier---in which every plan among the top ranks is
unbuildable---into one topped by a constructible plan, and doing so inside the
search rather than discarding afterward is worth +10.8\% of predicted
throughput over naively falling back to the first buildable candidate.

We hold our negative results to the same standard as our positive ones, with
the same reporting spec---number, caliber, attribution---and no softening. On
8$\times$H800, \sys's searched serving plan matches the strongest hand-tuned
configuration at a 1.0304 throughput ratio (argv-identical, three-round
median),
and on 2$\times$RTX4090 its searched training plan beats the strongest hand-tuned
baseline by +0.9\% and clears the acceptance bar---0.98$\times$ that
baseline---by +2.9\%.
On 8$\times$H800 training, however, the searched plan is a computable, honest
FAIL: it reaches 0.9338 of the best hand-tuned throughput (three-round
median), below the 0.98 bar, losing on a single overlap schedule flag---the
automatic and hand-tuned plans are the same \texttt{ep8} offload configuration
and differ only in whether overlap is enabled.
We pre-register these ranked predictions in a frozen, artifact-hashed
adjudication file before executing on the H800 testbed, and report every
outcome---including the ones that falsified our own predictions---as-is. The
result is a consistent stance: parity where the space is buildable, and
characterization where it is not.

This paper makes the following contributions:
\begin{itemize}
  \item \textbf{Realizability as a first-class search constraint.} A
  realizability predicate that reuses the emitter as a dry run (one source of
  truth, not a parallel rule base), turning an all-unbuildable top-N into a
  constructible top plan. We show the predicate is necessary but not
  sufficient, and pair it with a launch-probe fallback.
  \item \textbf{A closed-loop full-space optimizer for MoE.} A three-tier
  search over parallelism, schedule, and kernels, scored by dual-fidelity cost
  models, that emits a Megatron training stack and an SGLang serving stack
  from one plan, with measured realization overheads priced rather than
  forbidden.
  \item \textbf{A pre-registered, artifact-hashed evaluation.} We pre-register
  our ranked predictions in a frozen adjudication file before executing on the
  H800 testbed, and report the outcomes---including where our own predictions
  were falsified---as-is.\footnote{Code and every cited artifact are released
  with the paper so each \texttt{\%~SRC} pointer is externally checkable; see the
  ``Artifact availability'' note in \cref{sec:appendix} (release URL and license
  pending).}
  \item \textbf{Two-generation evidence, negatives included.} An evaluation on
  RTX4090 and H800 that reports a passing training result, serving parity, and
  a computable training FAIL, with every honest negative indexed rather than
  hidden.
\end{itemize}

The DualPipe plan above is our running example; we unfold it, and the
toolchain boundaries behind it, in \cref{sec:background}.
\Cref{sec:design,sec:impl} present \sys's design and implementation;
\cref{sec:eval} reports the two-generation evaluation; and \cref{sec:related}
places \sys against the closest automatic-parallel and kernel-generation
systems.

%% file: sections/02-background.tex
\section{Background and Motivation}
\label{sec:background}

Mixture-of-Experts (MoE) layers scale model capacity by routing each token to a
small subset of expert sub-networks, decoupling total parameter count from
per-token compute~\cite{shazeer2017moe,lepikhin2021gshard,fedus2022switch}.
Frontier models push this sparsity to extremes---DeepSeek-V3 activates 37B of
671B parameters, and Kimi~K2 activates 32B of 1.04T across 384 routed
experts---and this capacity gain comes with a corresponding increase in
infrastructure complexity~\cite{deepseek2024v3,kimi2025k2}. Crucially, one model
is deployed along several axes at once: DeepSeek-V3 runs training at
expert-parallel degree~64 but scales expert parallelism up to 320 for
decode~\cite{deepseek2024v3}, so no single parallelization plan is best across
training, prefill, and decode. Yet the systems that turn a plan into runnable
code are split into two disjoint stacks: training uses (semi-)automatic
parallelism in frameworks such as Megatron-LM~\cite{shoeybi_megatronlm}, while
large-scale serving relies on hand-configured expert parallelism in engines such
as SGLang~\cite{sglang2025largeep}. In both stacks cross-device expert
communication is a first-order cost---32--47\% of step time in production MoE
training~\cite{comet2025,megascalemoe2026}---so the plan, its communication
schedule, and the kernels that realize it are coupled and cannot be chosen in
isolation. \sys targets both stacks with one optimizer.

\subsection{A ranked plan you cannot build}
Consider the plan our optimizer ranks first for training our reference MoE on
eight H800 GPUs. Before running anything on the testbed, we pre-register the
ranked predictions in a frozen adjudication file and report the outcomes as-is.
The top-ranked training plan is \texttt{tp1/pp1/ep8/dualpipe+ov}, predicted at
24{,}712.9 tok/s/GPU under default cost constants and 27{,}758.2 tok/s/GPU once
the cost model is calibrated to H800 measurements.
By the analytic memory model the plan is comfortably feasible: it needs
46.2~GB per GPU against a 72~GB budget (80~GB HBM at a 0.9 usable fraction),
roughly 26~GB of headroom.

The plan is a mirage. It fails only when we try to construct it: the emit stage,
which lowers a plan to a concrete Megatron launch command, raises a
\texttt{NotImplementedError}---Megatron core~v0.18.0 has no DualPipe schedule,
and our emitter refuses to silently emit a command that ignores the plan's
\texttt{dualpipe} schedule.
The plan is therefore \emph{unrealizable}: it lives in the space the cost model
ranks but not in the space the toolchain can build. This is not a calibration
error---the better-calibrated model ranks the same plan \emph{higher}, not
lower---it is a realizability failure. Serving shows the same shape: the
pre-registered serve top-1 \texttt{tp1/pp1/ep4} (predicted 3985.2 tok/s) is
unrealizable-as-modeled because dp-attention requires \texttt{dp==ep} while eight
GPUs force \texttt{dp}=8, and the plan the hardware actually prefers,
\texttt{tp8}, never appears in the pre-registered top-5.
In both cases the optimum of the ranking sits outside the buildable set: a cost
model can be arbitrarily well calibrated and still rank plans that no downstream
tool can construct. This gap---between the space the cost model ranks and the
space the toolchain can build---is what \sys is built to close. It also explains
why realizability cannot be bolted on as a final check: the plans most worth
building are exactly the aggressive ones---novel schedules, maximal expert
parallelism---that stress the toolchain hardest.

\subsection{Eight toolchain boundary layers}
The DualPipe plan is one instance of a broader pattern. Across the training and
serving stacks we catalog eight distinct \emph{toolchain boundary layers} at
which a highly-ranked plan can fail to build (\cref{tab:boundaries}, B1--B8); one
of the eight, B3, is already cleared and superseded by B4, kept as an audit-trail
entry rather than counted as an active crack.
They fall into two categories. \emph{Tooling boundaries} (B1, B2, B3, B5) are
plan dimensions the emitter or launcher cannot map---a missing schedule, an
indivisible layer count, an unmapped optimizer flag, an enumeration constraint.
\emph{Cost-model boundaries} (B4, B6, B7, B8) are plans the model admits and the
hardware then rejects, and several of these survive even after the cost model is
corrected: fixing the expert-DP optimizer-shard accounting (B4) still leaves a
non-optimizer memory residual (B6) that over-admits plans the GPU cannot run.
We report each layer's predicate effect, by number, in \cref{sec:eval}.

\begin{table*}[t]
\centering
\caption{The eight toolchain boundary layers at which a highly-ranked plan can
fail to build. \emph{Layer} marks where in the toolchain the plan is rejected;
\cref{sec:eval} (RQ1) revisits each layer's predicate effect by number.}
\label{tab:boundaries}
\small
\begin{tabular}{@{}p{0.028\textwidth}>{\raggedright\arraybackslash}p{0.215\textwidth}>{\raggedright\arraybackslash}p{0.115\textwidth}>{\raggedright\arraybackslash}p{0.285\textwidth}>{\raggedright\arraybackslash}p{0.245\textwidth}@{}}
\toprule
\tblhead{\#} & \tblhead{Boundary} & \tblhead{Layer} & \tblhead{Evidence artifact} & \tblhead{Disposition} \\
\midrule
B1 & DualPipe schedule not emittable (Megatron core~v0.18.0) & tooling / emit & adjudication \texttt{rows[0-1]}, \texttt{NotImplementedError} verbatim & filed; \texttt{dualpipe\_startup\_div} left uncalibrated (P3 item~13) \\
B2 & Layer-count divisibility (27 layers vs.\ pipeline depth) & tooling / emit+launch & rows rank~3--5 + \texttt{pp2/pp4} launch-dead (torchrun traceback) & predicate coverage; enumerator unchanged \\
B3 & ZeRO stage unmapped $\rightarrow$ OOM (missing \texttt{distributed-optimizer} flag) & tooling / launch & \texttt{toolchain\_boundaries} verdict entry & cleared (amendment~1, \texttt{f8f5aa2}); superseded by B4 \\
B4 & Expert-DP optimizer-shard accounting (\texttt{ep==dp} over-admits) & cost-model / memory & amendment~1 new boundary class; ep-family OOM & fixed (amendment~2, \texttt{ea0859a}); residual $\rightarrow$ B6 \\
B5 & Serve \texttt{dp==ep} enumeration constraint (dp-attention) & tooling / emit & serve rows \texttt{ep4/ep2} unrealizable-as-modeled & predicate coverage \\
B6 & Non-optimizer memory residual ($\sim$27~GB; necessary not sufficient) & cost-model / launch & 47.0/57.8~GB admitted vs.\ $\sim$74~GiB OOM; scale-invariant & filed; launch-probe backstop retained \\
B7 & FSDP all-gather working buffer unmodeled & cost-model / launch & 69.32~GB predicted vs.\ $\sim$78--79~GiB OOM (0/6 emittable) & filed (P3 item~29) \\
B8 & Offload/overlap contention + $\alpha$-hiding domain failure & cost-model / schedule & penalty 0.0271 measured; domain-conditioned $-5.37\%$, no flip & penalty + domain-conditioning modeled; pattern item filed \\
\bottomrule
\end{tabular}
\vspace{2pt}
{\footnotesize\raggedright \emph{Environment-level boundaries excluded from this
catalog:} (i)~an \texttt{einops} hard dependency that blocks FSDP builds one
\texttt{pip install} short (amendments~7--8), and (ii)~a cuda-graph capture
conflict reproduced across sm\_89 and sm\_90 that forces
\texttt{-{}-disable-cuda-graph} (P3 item~20). Both are properties of the
execution environment rather than of the plan, and are handled outside the
search.\par}
\end{table*}

\subsection{Cracks, not bugs}
These eight layers are not a long tail of implementation bugs to be patched away
one by one. They are systematic cracks between ranking and construction: each is
a place where a plausible, well-scored plan meets a tool that cannot build it,
and several persist after the cost model itself is corrected. A post-hoc filter
that discards plans as they fail is not enough, because the failures are
structural and recur across models, hardware, and both stacks. Treating
realizability as a first-class search constraint---an emit dry-run the searcher
consults \emph{while} ranking, rather than after---is the design \sys develops in
\cref{sec:design}.

%% file: sections/03-design.tex
\section{Design}\label{sec:design}

\sys is a full-space optimizer for MoE workloads: from a single normalized
plan representation it searches jointly over parallelism strategy, communication
schedule, and generated kernel structure, scores candidates with a shared cost
model, and emits executable launch commands for both a training and a serving
backend. Its distinguishing commitment is that the search is
\emph{realizability-aware}: it ranks only over plans the toolchain can actually
build, closing the gap between the space the cost model ranks and the space the
toolchain can build. We describe the system in the normative present---what each
component does and why. \cref{sec:eval} traces how each principle below was
earned.

Four principles shape the design.

\smallskip\noindent\textbf{P1 (emit dry-run as the single source of truth).} The realizability
  predicate decides buildability by invoking the production emitter itself as a
  dry run, rather than re-encoding the toolchain's constraints in a parallel rule
  set. A candidate is realizable only if the emitter produces launch argv without
  raising; any guard the emitter grows---a new configuration assertion, a new
  divisibility precondition---is inherited by the predicate for free, so predicate
  and emitter can never drift. The dry run discards the argv and inspects only
  raise/no-raise, so it needs no real framework clone.

\smallskip\noindent\textbf{P2 (price, don't prohibit).} Costs that surface only at
  realization---the \emph{realization tax} of running a generated kernel through
  a patched execution stack, the contention penalty of overlapping communication
  on an offloaded optimizer path---are folded into the cost model as measured,
  multiplicative factors, not hard filters. The searcher sees the honest price of
  a plan and decides for itself whether to keep a kernel attached; a plan that
  cannot afford its tax simply loses on score rather than being excluded by fiat.
  This keeps a \emph{negative tax}, a generated kernel that beats its baseline,
  expressible on the same axis as a positive one.

\smallskip\noindent\textbf{P3 (the launch probe is a necessary-not-sufficient
backstop).} The predicate is a necessary condition for buildability, not a
  sufficient one: its memory gate can still admit plans that a real launch
  rejects. A live launch probe is therefore a mandatory, not optional, final step
  of the deployment protocol; the predicate narrows the buildable space but does
  not certify it. In this paper that probe is a rank-ordered walk-down we ran
  during the H800 campaign --- a manual/semi-automatic discipline over the bench
  harness, not yet a component the shipped \texttt{search()} invokes; folding the
  walk-and-launch loop into the searcher is future work (\cref{sec:discussion}).
  The necessary-not-sufficient principle is declared in the design, not appended
  after the fact.

\smallskip\noindent\textbf{P4 (domain-scoped calibration).} Every calibrated constant carries the
  domain it was fit over---the GPU, the tokens-per-expert range, the dtypes---and
  a query outside that domain is flagged \emph{extrapolated} rather than silently
  trusted. An uncalibrated cost model is bit-for-bit the factory model: the domain
  check is vacuously true when no domain is attached, so calibration is an overlay
  that adds hardware-specific knowledge without changing default behavior. A
  constant whose enabling assumption fails inside a domain is conditionalized away
  rather than extrapolated.
\smallskip

\begin{figure}
  \centering
  \input{figures/fig1-overview}
  \caption{\sys system overview. A normalized \emph{Plan} (parallelism layout,
  schedule, memory/sharding) is scored by a dual-fidelity cost model (analytic
  primary, simulator timeline), refined by a three-tier search
  (parallelism\,$\to$\,schedule\,$\to$\,kernel), and gated by the realizability
  predicate before dual-stack emission (Megatron for training, SGLang for serving)
  and Triton kernel codegen. The predicate (heavy border) is the first-class
  search constraint; a profiling/calibration loop feeds domain-scoped constants
  back into the cost model.}
  \label{fig:overview}
\end{figure}
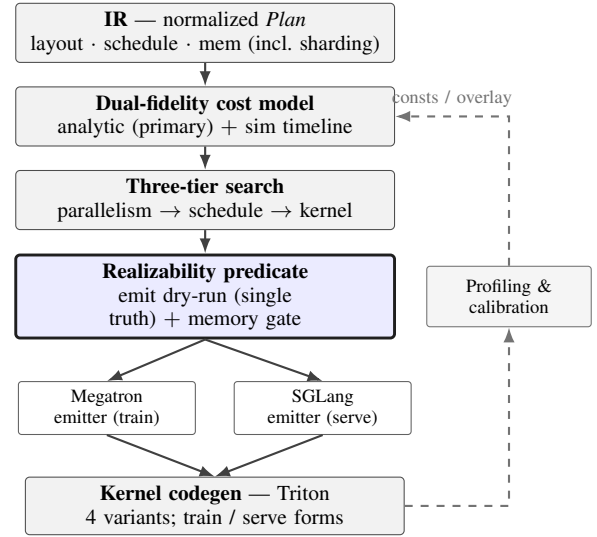

\cref{fig:overview} maps these principles onto the component chain, which we walk
in data-flow order: intermediate representation, cost model, search, and emitters.

\subsection{Intermediate representation}\label{sec:design-ir}

A \emph{Plan} is the single normalized object every stage consumes and produces
(\texttt{moefs/ir/plan.py}). It carries four sub-plans. The \emph{layout} fixes
the parallelism strategy---tensor, pipeline, and expert degrees
($\mathit{tp}$, $\mathit{pp}$, $\mathit{ep}$), an optional node limit, and a
redundant-expert count---with data parallelism derived from the world size. The
redundant-expert count is carried by the IR and the SGLang emitter but is
\emph{not} enumerated by the search: EPLB-style expert replication for
routing-skew mitigation is out of scope, so ``full-space'' spans the strategy,
schedule, and kernel axes, not load-balancing replication.
The
\emph{schedule} names the pipeline class (\texttt{1f1b}, \texttt{interleaved},
or \texttt{dualpipe}), whether all-to-all communication is overlapped, and the
communication chunking, SM budget, and placement that a timeline simulator needs.
The \emph{memory} plan carries the ZeRO stage, recompute flag, and a
\emph{sharding} dimension over $\{\texttt{zero1}, \texttt{zero1\_offload},
\texttt{fsdp}\}$; sharding is orthogonal to the layout and defaults to
\texttt{zero1}, which reproduces prior behavior bit-for-bit and keeps the offload
and fully-sharded regimes as opt-in memory levers.
Finally, an optional \emph{kernel} spec (\texttt{moefs/ir/kernel.py}) describes a
generated MoE kernel: its \emph{structural} dimensions are the fusion
boundaries---whether the token permute fuses into the grouped-GEMM prologue
(\texttt{fuse\_gather}) and whether the weighted combine fuses into its epilogue
(\texttt{fuse\_combine}), a $2\times2$ cross-product of variants---while its
\emph{numeric} dimensions (tile shape, warps, stages) form the autotuning space.
Making fusion a search variable, not a hand-written constant, is what places the
kernel inside the same plan the strategy and schedule live in. A reserved
\texttt{comm\_kind} field distinguishes stream-initiated from device-initiated
communication; the device-initiated value is refused at construction on hardware
where it cannot be validated, so an unmappable dimension fails loudly rather than
being silently emitted. This realizes the substrate P1 and P2 act on: one plan
carries every knob, so both buildability and price are decided against one object.

\subsection{Dual-fidelity cost model}\label{sec:design-cost}

Candidates are scored by two evaluators that share one constant set
(\texttt{moefs/cost/consts.py}) and produce the same throughput metric, so their
outputs are directly comparable. The \emph{analytic} model
(\texttt{moefs/cost/step\_time.py}) is a closed-form step-time estimate; it is the
primary scorer because it is cheap enough to sweep the full enumeration grid. The
\emph{simulator} (\texttt{moefs/cost/step\_time\_sim.py}, driven by
\texttt{moefs/sim/engine.py}) is a list-scheduling timeline that models the
overlap of compute and communication explicitly: a per-layer pipeline for the
1f1b and interleaved classes and a steady-state window for dualpipe, from which it
derives the achieved fraction of communication hidden behind compute. The
simulator is the higher-fidelity scorer used to refine schedules, where the
overlap that the analytic $\alpha$-table only approximates must be modeled
directly.

The realization tax lives here, as principle P2 prescribes. It is a set of lazy
constants---the per-leg kernel-patch overhead
(\texttt{kernel\_patch\_overhead\_frac\_serve} and \texttt{\_train}) and the
offload/overlap contention penalty (\texttt{offload\_overlap\_penalty\_frac})---each
defaulting to \texttt{None}. When unset, scoring is unchanged to the bit; when a
domain has been calibrated, the applicable factor scales predicted throughput
multiplicatively (throughput $\div\,(1+\text{frac})$), because the measured
quantity is a whole-stack throughput ratio rather than a decomposable time term.
A tax applies only when the plan would actually take the taxed path---an attached
kernel for the patch overhead, an offloaded-and-overlapped schedule for the
contention penalty---so the searcher prices exactly the plans that pay. Under P4,
each such constant is carried by an overlay keyed to a calibration domain
(\texttt{configs/calibrated/h800.yaml}): the overlay pairs the fitted values with
a \texttt{calibrated} block naming the GPU, the tokens-per-expert range, and the
dtypes it covers, and the model marks any query outside that range as
extrapolated. An uncalibrated model exposes no tax and behaves as the factory
default. The concrete tax magnitudes, and how they invert across GPU generations,
are reported in \cref{sec:eval}.

\subsection{Three-tier search and the realizability predicate}\label{sec:design-search}

Search proceeds in three tiers of increasing fidelity and cost
(\texttt{moefs/search/searcher.py}). The \emph{parallelism} tier enumerates
layouts---tensor degrees over the head-divisible subset of $\{1,2,4,8\}$,
pipeline and expert degrees over the admissible divisors, and, when enabled, the
sharding cross-product---and discards each candidate that fails a structural
divisibility check or the memory gate before scoring the survivors with the
analytic model. The \emph{schedule} tier takes the top-ranked layouts (a short
list of ten) and refines each over the schedule grid with the simulator,
substituting the refined schedule and score.
The \emph{kernel} tier then queries an offline autotuning report for the winning
kernel spec at each layout's expert shape, attaches it, re-scores with the
simulator, and keeps the attachment only when it wins---so a plan that cannot beat
its own unattached score keeps \texttt{kernel=None} rather than claiming a benefit
it does not have.

\begin{figure}
  \centering
  \input{figures/fig2-predicate}
  \caption{Realizability predicate flow, in code order
  (\texttt{moefs/search/realizability.py}). A candidate plan is checked by an
  \emph{emit dry-run} first---rejecting, with the verbatim exception, the boundary
  classes the emitter guards (dualpipe, layer-count indivisibility, pipeline/overlap
  preconditions, serving $\mathit{dp}\!=\!\mathit{ep}$)---then by a memory gate. The
  predicate yields a hard verdict (dashed bracket). Realization-tax pricing
  (``price, don't prohibit'') and the launch probe (a necessary-not-sufficient
  backstop that catches plans the memory gate over-admits) are the surrounding
  doctrine acting on plans the predicate admits.}
  \label{fig:predicate}
\end{figure}
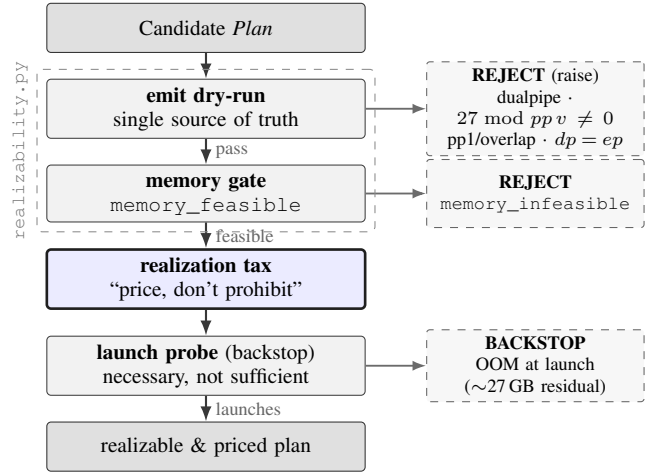

The realizability predicate is the search's first-class constraint
(\texttt{moefs/search/realizability.py}). Given a plan, it returns a verdict and,
on rejection, the verbatim triggering reason, so the pruned set is attributable
rather than a bare count. Its order is fixed: an \emph{emit dry-run} first, the
memory gate second (\cref{fig:predicate}). The dry run calls the training emitter
for a training phase and the serving emitter---with the real world size---for a
serving phase, and treats any \texttt{NotImplementedError} or \texttt{ValueError}
as unbuildable. Reusing the emitter as the oracle (P1) means the dry run covers,
without restating, the emit-side members of the eight toolchain boundary layers
of \cref{sec:background}: a pipeline class the backend cannot schedule, a layer
count that does not divide the pipeline, a serving layout whose data-parallel and
expert degrees disagree. The memory gate then calls the same feasibility check the
enumeration uses; the remaining boundaries are priced by the realization tax (P2)
or caught by the launch probe (P3) rather than pruned here.

The predicate is applied at two points. At the parallelism tier it prunes
unbuildable layouts before they are scored, so an unbuildable plan never enters the
ranking; the same predicate is threaded into schedule refinement as a consistency
guard, so refinement cannot substitute an unbuildable schedule (such as dualpipe)
back into an otherwise buildable layout. The result is an invariant: every plan in
the returned top-$k$ passes the predicate. The whole mechanism is opt-in---the
search defaults to an unconstrained mode that reproduces prior rankings
bit-for-bit, and the deployment pipeline requests the constrained mode
explicitly---so realizability constrains the space without perturbing the
oracle rankings that measure the cost model's taste. Crucially, the predicate is
declared necessary but not sufficient (P3): passing its memory gate does not
guarantee a launch, because the analytic accounting omits non-optimizer memory,
so a live launch probe backstops it. This realizes P1 and P3.

\subsection{Dual-stack emitters}\label{sec:design-emit}

The same Plan lowers to executable launch commands on two backends: a training
emitter targeting Megatron (\texttt{moefs/emit/megatron.py}) and a serving emitter
targeting SGLang (\texttt{moefs/emit/sglang.py}), together giving \sys its unified
training and serving reach. Each returns pure argv---element zero is the real
executable---with provenance recorded separately, and each command's argv is
fingerprinted by a hash so that a committed artifact can prove whether two
configurations truly emitted different commands. Because the same emitter code is
what the realizability predicate exercises as a dry run, the buildability the
search reasons about and the commands it finally produces are one artifact, never
two views that can disagree; this is P1 at the boundary of the system. The emitter
guard chain, the EP-within-TP realization semantics, and the kernel-patch
registration path are detailed in \cref{sec:impl}.


%% file: figures/fig1-overview.tex
\begin{tikzpicture}[
  font=\footnotesize,
  node distance=3.1mm,
  stage/.style={rectangle, rounded corners=1.6pt, draw=black!70, fill=black!5,
                align=center, inner sep=3pt, text width=48mm, minimum height=7.5mm},
  gate/.style={rectangle, rounded corners=1.6pt, draw=black!88, line width=1.1pt,
               fill=blue!8, align=center, inner sep=3pt, text width=48mm,
               minimum height=7.5mm},
  sub/.style={rectangle, rounded corners=1.2pt, draw=black!62, fill=white,
              align=center, inner sep=2pt, text width=22mm, minimum height=7mm,
              font=\scriptsize},
  side/.style={rectangle, rounded corners=1.2pt, draw=black!55, fill=black!4,
               align=center, inner sep=2.5pt, text width=20mm, minimum height=8mm,
               font=\scriptsize},
  flow/.style={-{Latex[length=2mm]}, draw=black!75, line width=0.8pt},
  fb/.style={-{Latex[length=2mm]}, draw=black!55, line width=0.8pt, dashed},
]

\node[stage] (ir)
  {\textbf{IR} --- normalized \emph{Plan}\\layout $\cdot$ schedule $\cdot$ mem (incl.\ sharding)};
\node[stage, below=of ir] (cost)
  {\textbf{Dual-fidelity cost model}\\analytic (primary) $+$ sim timeline};
\node[stage, below=of cost] (search)
  {\textbf{Three-tier search}\\parallelism $\to$ schedule $\to$ kernel};
\node[gate, below=of search] (real)
  {\textbf{Realizability predicate}\\emit dry-run (single truth) $+$ memory gate};

\node[sub, below=5.5mm of real.south, xshift=-13mm] (mega) {Megatron\\emitter (train)};
\node[sub, right=5mm of mega] (sgl) {SGLang\\emitter (serve)};

\node[stage, below=5.5mm of $(mega.south)!0.5!(sgl.south)$, text width=48mm] (kern)
  {\textbf{Kernel codegen} --- Triton\\4 variants; train / serve forms};

\node[side, right=4mm of real] (prof) {Profiling \&\\calibration};

\draw[flow] (ir) -- (cost);
\draw[flow] (cost) -- (search);
\draw[flow] (search) -- (real);
\draw[flow] (real.south) -- (mega.north);
\draw[flow] (real.south) -- (sgl.north);
\draw[flow] (mega.south) -- (kern.north);
\draw[flow] (sgl.south) -- (kern.north);

\draw[fb] (kern.east) -| (prof.south);
\draw[fb] (prof.north) |- node[pos=0.75, above, font=\scriptsize, black!55]
        {consts / overlay} (cost.east);

\end{tikzpicture}

%% file: figures/fig2-predicate.tex
\begin{tikzpicture}[
  font=\footnotesize,
  node distance=3.4mm,
  pbox/.style={rectangle, rounded corners=1.6pt, draw=black!75, fill=black!5,
               align=center, inner sep=3pt, text width=40mm, minimum height=7.5mm},
  price/.style={rectangle, rounded corners=1.6pt, draw=black!88, line width=1.0pt,
                fill=blue!8, align=center, inner sep=3pt, text width=40mm,
                minimum height=7.5mm},
  term/.style={rectangle, rounded corners=1.6pt, draw=black!80, fill=black!12,
               align=center, inner sep=3pt, text width=40mm, minimum height=6.5mm},
  rej/.style={rectangle, rounded corners=1.2pt, draw=black!60, dashed, fill=black!3,
              align=center, inner sep=2.5pt, text width=27mm, minimum height=9mm,
              font=\scriptsize},
  flow/.style={-{Latex[length=2mm]}, draw=black!78, line width=0.8pt},
  rflow/.style={-{Latex[length=1.8mm]}, draw=black!58, line width=0.7pt},
]

\node[term] (cand) {Candidate \emph{Plan}};
\node[pbox, below=of cand] (dry) {\textbf{emit dry-run}\\single source of truth};
\node[pbox, below=of dry] (mem) {\textbf{memory gate}\\\texttt{memory\_feasible}};
\node[price, below=of mem] (tax) {\textbf{realization tax}\\``price, don't prohibit''};
\node[pbox, below=of tax] (probe) {\textbf{launch probe} (backstop)\\necessary, not sufficient};
\node[term, below=of probe] (ok) {realizable \& priced plan};

\node[rej, right=8mm of dry] (r1)
  {\textbf{REJECT} (raise)\\dualpipe $\cdot$ $27\bmod pp\,v\neq0$\\pp1/overlap $\cdot$ $dp\!=\!ep$};
\node[rej, right=8mm of mem] (r2) {\textbf{REJECT}\\\texttt{memory\_infeasible}};
\node[rej, right=8mm of probe] (r3) {\textbf{BACKSTOP}\\OOM at launch\\($\sim$27\,GB residual)};

\draw[flow] (cand) -- (dry);
\draw[flow] (dry) -- node[right=0pt, font=\scriptsize, black!60] {pass} (mem);
\draw[flow] (mem) -- node[right=0pt, font=\scriptsize, black!60] {feasible} (tax);
\draw[flow] (tax) -- (probe);
\draw[flow] (probe) -- node[right=0pt, font=\scriptsize, black!60] {launches} (ok);

\draw[rflow] (dry.east) -- (r1.west);
\draw[rflow] (mem.east) -- (r2.west);
\draw[rflow] (probe.east) -- (r3.west);

\node[fit=(dry)(mem), draw=black!45, dashed, rounded corners=2pt,
      inner sep=3.5pt, label={[font=\scriptsize, black!55, rotate=90,
      anchor=south]west:{\texttt{realizability.py}}}] (box) {};

\end{tikzpicture}

%% file: sections/04-implementation.tex
\section{Implementation}\label{sec:impl}

\sys{} is implemented in Python over PyTorch and Triton, targeting the pinned Megatron-LM (\texttt{core\_v0.18.0}) and SGLang (\texttt{0.5.15}) toolchains. This section describes the three components that turn a ranked \texttt{Plan} into an executable artifact, and the measurement infrastructure that underwrites every number in \cref{sec:eval}. We report the engineering detail that exposes a discipline or a semantic mapping, and omit installation mechanics.

\subsection{Emitters}

The emitters lower a \texttt{Plan} to pure executable \texttt{argv}: \texttt{emit\_megatron\_train} returns a \texttt{torchrun} command and \texttt{emit\_sglang\_serve} a \texttt{python -m sglang.launch\_server} command, each safe to hand directly to \texttt{subprocess.run}. Provenance is produced separately so that no comment element ever contaminates \texttt{argv[0]}.

Parallelism widths are mapped onto each toolchain's real argument surface, verified against the pinned commits rather than documentation. SGLang realizes expert parallelism within the tensor-parallel world (EP-within-TP: \texttt{ServerArgs} asserts \texttt{ep\_size*moe\_dp\_size==tp\_size}), whereas \sys's \texttt{Plan} uses Megatron-style EP-within-DP. A Plan with tensor width $T$ and expert width $E$ therefore does not realize as the naive \texttt{-{}-tp-size $T$ -{}-ep-size $E$}, which fails \texttt{ServerArgs} validation for every $E>1$; it realizes as a world of $T{\cdot}E$ ranks, \texttt{-{}-tp-size $T{\cdot}E$ -{}-ep-size $E$ -{}-enable-dp-attention -{}-dp-size $E$}: attention replicated data-parallel over $E$ groups of TP-$T$, experts sharded $E$-way.

The emitters are the single source of truth for realizability (\cref{sec:design}): any Plan dimension the pinned toolchain cannot express raises loudly rather than degrading to a silently-equivalent command. A \texttt{dualpipe} schedule raises \texttt{NotImplementedError} (Megatron \texttt{core\_v0.18.0} has no DualPipe scheduler); full recompute combined with expert-parallel communication overlap, and \texttt{pp>1} under a \texttt{1f1b} schedule with that overlap, each raise \texttt{ValueError} against the exact \texttt{transformer\_config.py} assertions they would trip; on the serve side, a Plan whose data-parallel width disagrees with $E$ raises \texttt{NotImplementedError} rather than dropping the extra axis. This guard chain refuses to emit a command that pretends to have realized a dimension it dropped.

One caveat on reading a rejection: because the predicate \emph{is} the emitter (\cref{sec:design}), an ``unrealizable'' verdict can be \emph{toolchain-inherent} (Megatron \texttt{core\_v0.18.0} has no DualPipe scheduler, independent of our IR) or an artifact of what the \texttt{Plan} IR can \emph{express} (e.g.\ whether \texttt{dp}${\neq}$\texttt{ep} dp-attention is truly impossible in SGLang or merely inexpressible in our EP-within-DP schema). The DualPipe headline is the former; the two are indistinguishable from the predicate's boolean alone, so IR-expressiveness over-rejection is a distinct, unquantified failure mode from the memory gate's disclosed under-rejection.

\subsection{Kernel generation}

A \texttt{KernelSpec} separates structural dimensions---the fusion boundaries \texttt{fuse\_gather} (token permute folded into the GEMM prologue) and \texttt{fuse\_combine} (weighted combine folded into the epilogue)---from numeric dimensions (tile sizes, warps, stages). The fusion boundaries are \texttt{tl.constexpr} switches, so the four combinations (unfused, gather-only, combine-only, fully fused) compile into structurally distinct kernels with different instruction and memory streams; the numeric dimensions are the autotune space. Kernel structure is thus a search variable, not a fixed hand-written kernel with tunable constants.

Training and decode share a single generated kernel rather than two implementations. The kernel body reads each expert's row segment through two independent offset pointers: the training path passes a monotone \texttt{[E+1]} offset array (with \texttt{hi = offs[1:]}), while the masked-decode path passes equal-stride capacity offsets plus valid-count offsets, giving the valid prefix inside each capacity segment. The body and mask logic are identical, so no second kernel is produced.

Each variant is checked against a reference over a shape grid and beyond-grid probes (non-aligned $K/N$, zero-token experts), with mutation testing confirming the store-mask and atomic row-mask are load-bearing.

\subsection{Engine registration}

Both generated engines register into unmodified production stacks without patching their source. Registration uses a \texttt{sitecustomize} trampoline placed on \texttt{PYTHONPATH}: the interpreter auto-imports it at startup, so every process in the engine's subprocess tree---for SGLang, the scheduler, detokenizer, and TP workers that \texttt{launch\_server} spawns from the inherited environment---applies the patch. The trampoline is import-light and lazy; \texttt{apply()} is idempotent and its own failure is loud, yet a failed patch never aborts interpreter startup.

For serving, \texttt{apply()} monkeypatches \texttt{FusedMoE.forward\_impl}. This seam sits above the single-GEMM primitive (whose EP dispatch layer we would otherwise reimplement) and below \texttt{FusedMoE.forward} (whose cuda-graph branching we would otherwise have to re-validate); a quantized layer hits \texttt{NotImplementedError} rather than a silent fallback.

For training, the generated kernels are wrapped in a \texttt{torch.autograd.Function} (\texttt{SequentialMLPFunction}) that replaces Megatron's \texttt{SequentialMLP.forward}: the forward composes the two GEMMs, and the backward composes the dual GEMMs from the same generator in transposed layout, delegating activation gradients to autograd and per-expert weight gradients to batched \texttt{torch.mm} (a disclosed v1 boundary). Both legs are proven numerically: SGLang per-layer relative difference 0.32\%/0.39\% against a 1\% threshold, and Megatron 15-step loss alignment 0.5313\% with backward parity 6/6.
The registration modules live at \texttt{moefs/kgen/\{sglang,megatron\}\_patch.py}, with the emit-side artifact generator at \texttt{moefs/emit/kernel\_patch.py}.

\subsection{Measurement infrastructure}

Every number in \cref{sec:eval} rests on one of two infrastructures of different strength: an automated phase gate that anchors the local $2{\times}$RTX4090 results, and the append-only, hash-locked H800 adjudication that anchors the H800 results. The local gate does \emph{not} judge the H800 artifacts (\cref{sec:appendix}): the headline H800 FAIL and serving parity are backstopped by the adjudication, not the exit-code gate.

The phase gate is a pure function over committed on-disk bench artifacts---no GPU, no subprocess---so it recomputes identically for any reader. It asserts two success gates, train and serve, each requiring the searched top-1 throughput to reach $0.98{\times}$ the best hand-tuned baseline; and two staleness invariants (\texttt{searched\_top1\_matches\_fresh\_search}, one per leg) that re-run the search and fail the exit code if the committed top-1 no longer matches a fresh search. Both the gates and the staleness checks feed a single exit code, so a stale or regressed artifact cannot pass silently.
The machine-readable open-issue ledger \texttt{P3\_ENTRIES} carries 31 entries, each a structured record of a known limitation rather than prose.

The preregistered H800 adjudication is append-only: the ranking artifact committed before any H800 measurement is never rewritten, and its ten revisions are recorded as \texttt{amendment\_1} through \texttt{amendment\_10}, each byte-round-trip-verified against prior content and locked by a hash invariant in the test suite. This is the base on which \cref{sec:eval} reports both the calibrated matches and the honest negatives; we do not expand those results here.

%% file: sections/05-evaluation.tex
\section{Evaluation}\label{sec:eval}

We evaluate \sys around five questions. RQ1: does the realizability
predicate change which plans survive search? RQ2: does the search win end to
end under the paper's success criterion? RQ3: do the preregistered predictions
adjudicate honestly? RQ4: how do the generated kernels and the realization tax
behave across GPU generations? RQ5: does the search retain discriminative power
under ablation? All measurements come from two testbeds --- a local
$2{\times}$RTX4090 node (sm\_89) and an $8{\times}$H800 node (sm\_90, NVLink);
machine specifications, round counts, and protocol calibers are in
\cref{sec:appendix}. Honest negatives are reported with the same specification
as positive results --- number, caliber, attribution, no defensive adverbs ---
and every negative is indexed in \cref{tab:negatives}.

\subsection{RQ1: Realizability changes the surviving frontier}

The realizability predicate is not a cosmetic filter. On the $8{\times}$H800
testbed the well-calibrated cost model's top-5 training frontier is entirely
unbuildable: the union of six frontier plans yields zero measurable points
(\texttt{frontier\_union\_plan\_count} $6$, \texttt{frontier\_measured\_count}
$0$), and four adjudicated neighborhood attempts likewise produced zero measured
points, so no measured training throughput existed at adjudication time
(\texttt{no\_measured\_train\_point} true) under \emph{both} the default and the
calibrated constants.
The frontier's top-1 is the DualPipe plan whose unbuildability is the running
example of \cref{sec:background}; we reference its conclusion here without
re-narrating it. This is the load-bearing observation of the paper: the top of
the ranking space --- the plan the calibrated cost model likes best --- lies
outside the space the toolchain can build, and calibration only makes the model
rank it \emph{higher} ($27{,}758.2$ predicted), so the gap is one of
realizability, not accuracy.
That this is the top-5 of a \emph{well-calibrated} model matters: fixing
prediction accuracy, which we do in RQ3, leaves the frontier $0/6$ buildable, so
realizability is a distinct axis rather than a symptom of a weak cost model. The
objection that the predicate is ``merely engineering validation'' does not
survive the frontier being entirely unbuildable for the model's own best
guesses.

The eight toolchain boundary layers catalogued in \cref{tab:boundaries} are
disposed of by three predicate mechanisms rather than one rule
base.\footnote{Two further environment-level boundaries --- the \texttt{einops}
hard-block and the cuda-graph capture conflict --- are excluded from the count
of eight; they are container/runtime conditions rather than plan-structure
boundaries.} The \emph{emit dry-run} rejects the tooling boundaries at emit
time: layer~1 (DualPipe, \texttt{NotImplementedError}), the interleaved half of
layer~2 (layer-count divisibility, $27 \bmod pp\,v \neq 0$; its 1f1b half
surfaces only at launch --- \cref{tab:boundaries} lists B2 as emit+launch), and
layer~5 (serve $dp{=}ep$ for dp-attention). Layer~3 (\texttt{zero\_stage} unmapped $\to$ OOM) was
\emph{cleared} once the emitter learned to map \texttt{-{}-use-distributed-optimizer};
its successor is layer~4 (expert-DP optimizer sharding), which the corrected
\emph{memory gate} now rules. Layers~6--8 --- the ${\sim}27$~GB non-optimizer
residual, the unmodeled FSDP all-gather buffer, and offload/overlap contention
--- are the ones the cost model still over-admits, and they are backstopped by
the \emph{launch probe}. Each layer's disposition is a predicate mechanism, not
a hand-written special case.

Given the predicate, search re-refines the constraint inside the buildable
domain rather than filtering after the fact. Its buildable training top-1
(\texttt{tp1/pp1/ep8/1f1b+ov}, predicted $26{,}174.9$ tok/s/GPU) beats the naive
alternative --- discard the unbuildable ranking and retreat to the first
buildable base candidate (\texttt{tp1/pp2/ep4/1f1b}, $23{,}614.5$) --- by
$+10.8\%$ ($26174.91/23614.55-1 = +10.84\%$).
The gain comes from folding the constraint \emph{into} search --- the buildable
top-1 recovers the ep8 overlap it would otherwise lose --- not from a post-hoc
filter over a ranking whose head is already dead. Equivalently, treating
unbuildability as a hard search constraint rather than a post-filter is worth
$+10.8\%$ of predicted throughput on this frontier alone.

The predicate is cheap enough to run inside search at scale. On the paper's own
published 2048-GPU oracle (256~nodes~$\times$~8~H800, DeepSeek-V3), enumeration
yields $4{,}820$ candidates; the emit dry-run plus memory gate over \emph{every}
one of them takes $22.3$~ms ($4.6$~$\mu$s per candidate), and the full three-tier
search with the realizability constraint on completes in $70.2$~ms (median of
three CPU-only rounds); the local $2{\times}$4090 grid finishes in $13.9$~ms.
CPU-only and needing no framework clone, realizability-checking every candidate
is not a cost the search must budget around.

\emph{Necessary but not sufficient.} The memory gate is a nontrivial negative
result in its own right, and it falsified our own accounting. After we corrected
the expert-DP optimizer sharding, ep4 and ep8 are still judged feasible ($57.8$
and $47.0$~GB within the $72$~GB budget), yet the real machine OOMs at
${\sim}74$~GiB --- a ${\sim}27$~GB non-optimizer residual (activation, bf16
gradient copies, framework buffers), since the expert optimizer state itself is
ep-invariant at $21.59$~GB ($22.93$~GB total optimizer) and cannot account for
it.
The same correction \emph{does} correctly flip ep2 ($79.39 > 72$~GB), matching a
real ep2 OOM --- evidence that the fix is right, not that the model is complete.
The launch probe then earned its keep four times in one sprint (we count the
four P4b.6 falsifications; the two earlier ep8-OOM backstops of P4b.5 are
separate), twice by falsifying \emph{our own} predictions: (i) offload was
predicted unbuildable, blocked by a TE dependency, but ran CLEAN $5/5$ steps
because the hybrid device optimizer bypasses that path; (ii) FSDP was predicted
launch-probable ($69.32 < 79.18$~GB) but was blocked pre-build by an
\texttt{assert HAVE\_EINOPS}; (iii) once \texttt{einops} was installed the same
FSDP $69.32$~GB prediction was directly falsified --- all six configurations
OOM at ${\sim}78$--$79$~GiB, ep1 dying before activations even allocate; and
(iv) the v3c hypothesis that removing the $\alpha$-hiding benefit would flip the
training verdict was itself falsified by v3d (RQ2 below).
The residual is scale-invariant (same peak at gbs $8$ and $16$), so the launch
probe is empirically required, not an optional safety net: the predicate admitted
a plan the machine then refused, and the backstop is where that modeling error is
caught, not hidden.

\subsection{RQ2: End-to-end panorama, including the leg we lose}

\cref{tab:panorama} reports the success criterion --- automatic $\geq
\max(\text{hand-tuned}) \times 0.98$, raw-float --- across two machines and both
phases. Three of four cells PASS and one FAILs; we report the FAIL first. The
asymmetry is deliberate: the FAIL is the leg we could most easily have hidden by
re-tuning, so putting it first is the strongest evidence that the pipeline is
not tuned to pass.

On $8{\times}$H800 training the terminal result is a FAIL at ratio $0.9338$
(automatic $12{,}059.0$ vs.\ the best hand-tuned $12{,}913.2$ tok/s/GPU;
threshold $12{,}655.0$).
We characterize this negative at three layers. \emph{Kind}: it is a
\emph{computable} honest FAIL, not an unbuildable frontier. The training leg was
an honest-negative for two prior versions --- v1 could run $0/7$ candidates (six
CUDA OOM, one model-infeasible dp8 predicted at ${\sim}120.8$~GB), and v2
produced $0$ probe-clean configurations over twelve probes --- and became
measurable for the first time only after the sharding dimension was added (v3,
ratio $0.9714$).
So unlike the preregistered honest-negatives, this prediction could finally be
tested, and it fails the $0.98$ rule. The distinction matters
for scoring: an unbuildable frontier can only be an honest-negative, whereas
a computable FAIL is a real loss on a real machine and we score it as one. \emph{Locus}: it is a cost-model boundary
(layer~8 of \cref{tab:boundaries}, offload/overlap contention under
$\alpha$-hiding domain failure), not a tooling boundary; automatic and
hand-tuned run the identical ep8 offload configuration and differ only by one
overlap flag --- we lose on a single schedule flag.
\emph{Robustness}: three estimators --- $0.9338$ formal 3-round median,
$0.9690$ confirmation round, $0.9517$ four-round merged --- all fall below
$0.98$. We define the noise band we disclose: the automatic $+$ov leg runs on a
copy-bound offload path (GPU utilization $0$--$26\%$, host--device copy
dominated) whose per-round throughput carries a ${\sim}{\pm}5\%$ run-to-run
variance. Its three formal rounds ($11{,}354.6$, $12{,}059.0$, $12{,}983.6$
tok/s/GPU) span a ${\sim}13.5\%$ range, wider than the compute-bound hand-tuned
no-ov leg ($13{,}403.6$, $12{,}647.3$, $12{,}913.2$; ${\sim}5.9\%$). The verdict
therefore rests on the preregistered rule --- the $0.98$ threshold on the
raw-float 3-round median --- not on a claim that every estimator clears the
noise: the confirmation round ($0.9690$) sits only ${\sim}1.1$~pp below the bar,
inside this band, which we say plainly rather than call the FAIL comfortably
robust. With $N{=}3$ we run no formal significance test: against a preregistered
raw-float rule, at this sample size a test would be formality, not evidence.

We had a choice and declined to tune the model to a PASS (the reusable discipline
is \cref{fig:checklist}). Four elements make this concrete:
(1) the calibration-domain argument precedes the verdict, not the reverse;
(2) the overlap penalty ($0.027086$) is the mean of two 3-round-median ov/no-ov
pairs ($0.0294$, $0.0248$), a measured anchor, not a value inflated to force a
flip;
(3) the $0.98$ rule is applied to raw floats and all three estimators are below
it; and (4) both sides were measured in the same session over three fresh
rounds. Our own v3c hypothesis --- that removing the $\alpha$-hiding benefit
would flip the verdict --- was then falsified by v3d: domain-conditioned removal
moved the $+$ov prediction from $25{,}484.6$ to $24{,}116.1$ ($-5.37\%$) and
narrowed the prediction gap from $+8.1\%$ to $+2.28\%$, yet the verdict did not
flip; the residual cause (a2a-vs-all-gather pattern asymmetry, $+5.05\%$) lay
beyond the authorized surface, so we exited terminal rather than keep editing
the model until it passed.
Because the amendment log is append-only, this reversal of our own hypothesis
is visible in the record rather than quietly dropped.

The three PASS cells are: local training, searched top-1 $38{,}028.3$ tok/s vs.\
the strongest baseline \texttt{megatron\_default\_dp2} $37{,}703.4$ (threshold
$36{,}949.3$), a $+2.9\%$ margin;
local serving, searched top-1 (\texttt{tp2}) $3{,}510.9$ vs.\
\texttt{handtuned\_tp2} $3{,}549.9$, ratio $0.9890 \geq 0.98$;
and H800 serving, automatic (\texttt{tp8}) $12{,}274.6$ agg.\ vs.\ hand-tuned
$11{,}912.5$, ratio $1.0304$.
The H800 serve PASS was itself earned: v1 was an honest FAIL at $0.8765$ because
the priced model honestly ranked ep8 while the hardware favored \texttt{tp8},
and the automatic side used its losing honest choice rather than fabricate a
win; the world-8 recalibration then cleared the tension to $1.0304$.
The four verdicts are \emph{not} same-source, and the table note says so: both
serving cells are argv-identical (byte-identical argv on both sides --- local
serve emits one command hash for searched and hand-tuned alike), so the H800
$1.0304 > 1$ is a $3.0\%$ same-configuration run-to-run noise band, not the
automatic side ``beating'' the hand-tuned side, while the local training
$+2.9\%$ is a margin against the $0.98{\times}$baseline threshold (only $+0.86\%$
against the baseline itself). Round-to-round spread is tightest on the PASS
cells --- local training ${\sim}0.4\%$/${\sim}0.3\%$, local serving
${\sim}1.1\%$/${\sim}1.1\%$, and H800 serving ${\sim}1.9\%$/${\sim}3.1\%$ over
three rounds (searched/hand-tuned) --- so all four cells carry the same
multi-round disclosure as the FAIL, and the noisiest leg is the FAIL's
copy-bound offload path, not a PASS margin. We flag these semantics rather than
read the cells as homogeneous wins.

\subsection{RQ3: Preregistered adjudication}

Predictions were committed and hashed before any H800 measurement; the
adjudicator is a pure consumer that copies predicted numbers field-by-field from
the frozen artifact, and revisions are append-only amendments, never rewrites.
\cref{tab:adjudication} shows the three verdict branches across two phases and
two constant sources. The branches distinguish two kinds of negative: a
\emph{tooling boundary} (the plan cannot be built) is not a
\emph{prediction error} (the plan is built but mispredicted); the training
prediction, unbuildable on this toolchain, is neither confirmed nor falsified
but simply untestable here.
The middle branch, WEAK (frontier or band agreement short of a top-1 match), is a
reserved tier no cell lands in as its \emph{final} label here: serve$\times$default
fails weak validation too (its realizable entries' m/p $0.285$--$0.323$ fall
outside any reasonable band), so that row is honest-negative, and
serve$\times$calibrated is a STRONG top-1 match. We keep WEAK in the discipline
(\cref{fig:checklist}) for studies that land there; ours converge to STRONG or
honest-negative.

The strong-validation story is serving under calibration. Under the default
constants --- a pure cross-hardware extrapolation that never saw an H800
measurement --- the three buildable serve plans have measured/predicted (m/p)
ratios of $0.285$, $0.301$, and $0.323$ (the model over-predicts throughput by
$3.1$--$3.5\times$) and the ranking is wrong: the preregistered top-1
(\texttt{ep4}) is unrealizable-as-modeled and the actual top-1 (\texttt{tp8}) is
not even in the preregistered top-5.
After calibration --- fused-curve autotuning plus a dp-attention bandwidth and
latency overlay --- the ratios converge to $0.738$, $0.749$, and $0.848$
(predicted/measured narrowed to $1.2$--$1.4\times$), the predicted top-1 matches
the measured top-1 (\texttt{tp8}), and the head ordering
$\texttt{tp8} > \texttt{tp4} > \texttt{ep8}$ is reproduced by the hardware. This
hits a criterion written down in advance in the run-book: if the default ratios
are markedly worse and calibration returns them to a reasonable band, the
calibration domain is hardware-dependent --- a finding, not a failure. The
domain scoping is concrete: the $\alpha$-hiding benefit is conditioned out in the
offload domain, where the compute-dominance premise fails (GPU utilization
$0$--$26\%$), which is exactly the hardware dependence the criterion predicted.
The calibration and the adjudication draw on \emph{different} measurements on the
same node and model: the constants are fit from microbenchmarks and profiling
(fused-kernel autotuning, a world-8 dp-attention bandwidth/latency overlay), not
from the end-to-end serving throughputs they then adjudicate, so the match is not
an in-sample tautology. It is, however, \emph{within}-hardware and
\emph{within}-model, with no cross-workload holdout on the serving side ---
unlike the RTX4090 training calibration, which holds one configuration out and
predicts it to $2.6\%$ (a threat we log in \cref{sec:discussion}).
Two guardrails are built in: the fine ordering sits inside inter-round noise (range
\texttt{tp4}~$28.0\%$, \texttt{tp2}~$17.4\%$, \texttt{ep8}~$12.6\%$ vs.\
\texttt{tp8}~$5.2\%$), and the magnitude is not closed --- measured throughput
still trails the calibrated prediction by $15$--$26\%$.

Serving is argv-identical on both machines, so a reader might discount it as
parity rather than a win. Three facts answer this. (a) The local serve top-1 was
once \emph{mis}-ranked: the earlier searched top-1 (\texttt{ep2} @$2{,}925.7$
pre-P3a, @$2{,}886.7$ pre-P3b) were both inferior to \texttt{handtuned\_tp2}
$3{,}549.9$, and only the P3b model fix moved the top-1 to \texttt{tp2}
@$3{,}510.9$ --- the correct answer was earned, and the history is on record.
(b) On H800 the default constants sort the plans entirely wrong, and calibration
recovering the ranking is a nontrivial prediction achievement, not a tautology.
(c) The training side is \emph{not} parity at all: local training is a genuine
win and the H800 training result is an honest negative. This follows the
oracle-suite tradition where the validation is that first-principles search
rediscovers the hand-tuned optimum --- and the one leg where it cannot, we
report as a loss.

\subsection{RQ4: Kernels and the realization tax across generations}

\begin{figure}[t]
\centering
\includegraphics[width=\columnwidth]{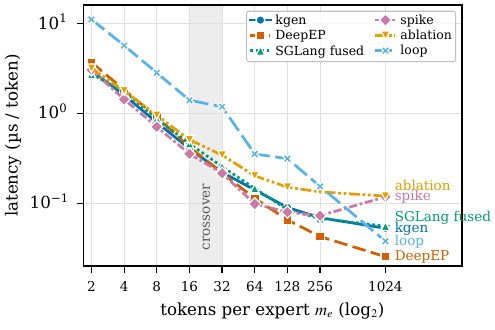}
\caption{Per-token latency, generated kernels vs.\ DeepEP full chain
($8{\times}$H800, NV8). \emph{Caliber box:} $y$ is per-token
$\mu$s/token (\emph{not} raw ms). kgen $=$ single-card pure expert compute;
DeepEP $=$ v1.2.1 (NVSHMEM) EP8 full chain (cross-rank dispatch $+$ expert GEMM
$+$ combine, MAX wall clock) --- \texttt{apples\_to\_apples=false}; the shared
denominator is global tokens $= me{\cdot}E/\text{top\_k}$. Shape
$E{=}64,K{=}2048,N{=}1408,\text{top\_k}{=}2$; protocol CUDA events, warmup $10$,
$me{<}16$ reps $200$ else $50$, 3-round median-of-medians. $me{=}512$ is absent
in the artifact (do not interpolate); crossover band $me{=}16$--$32$ shaded.
DeepEP V2 (2.1.0) is an honest-negative build failure (not plotted).}
\label{fig:crossover}
\end{figure}

\cref{fig:crossover} plots per-token latency for the generated kernels against a
DeepEP~\cite{deepseek2025deepep} full-chain baseline on $8{\times}$H800. At small expert multiplicity the
generated single-card kernel (the winner of the joint structure-and-numeric
autotune search) wins point-by-point ($me{=}2$ $+24.0\%$, $me{=}4$
$+14.6\%$, $me{=}8$ $+14.6\%$, $me{=}16$ $+4.2\%$), because it pays no cross-card
communication; a crossover band sits at $me{=}16$--$32$ (at $me{=}32$ DeepEP
edges ahead by $2.0\%$, inside measurement noise); and at large $me$ DeepEP wins,
by $19.4\%$ at $me{=}64$ up to $51.7\%$ at $me{=}1024$, as its amortized
all-to-all is outweighed by the growing expert compute it overlaps.
We do not claim to beat DeepEP; the large-$me$ regime is a loss and we report
it. The comparison is not apples-to-apples --- DeepEP times the cross-card
all-to-all dispatch, expert GEMM, and combine as a global wall clock, while the
generated kernel times single-card pure expert compute --- and the shared
per-token denominator (global tokens $= me{\cdot}E/\text{top\_k}$) is the only
axis on which the crossover is meaningful (caption caliber box). The python
SequentialMLP loop line is a different regime (fixed overhead amortizing to a
${\sim}1.2$~ms plateau for $me{\geq}32$), plotted for completeness. The reading
is a regime split: at the small expert counts typical of dp-attention serving
the generated kernel is competitive, while DeepEP remains the right tool for
wide-EP training at large $me$. Our DeepEP negative claim is
scoped to V2: DeepEP V2 (2.1.0, NCCL-Gin) was an honest-negative build failure
(three root causes; \texttt{cudaMemcpyBatchAsync} requires CUDA${\geq}12.8$,
unsolvable inside our red line), so the plotted baseline is V1 (1.2.1, NVSHMEM).

The realization tax --- \texttt{overhead\_frac} $= \text{unpatched}/\text{patched}
- 1$, applied multiplicatively as throughput $\div (1{+}\text{frac})$ --- flips
in both sign and magnitude across generations (\cref{tab:tax}). On serving the
tax rises from $+41.7\%$ on RTX4090 to $+123.6\%$ on H800: the H800's native
\texttt{fused\_moe} is fast enough that the dual-launch and no-cuda-graph
penalties are amplified, and every serve candidate is rejected on both machines.
On H800 the untaxed searcher attaches the kernel $5/5$ (top-1 \texttt{tp8}
@$1{,}614.2$ tok/s), but the taxed searcher rejects $0/5$ and falls back to the
loop curve (\texttt{ep8} @$1{,}290.8$), since candidate fused gains of
$13.3$--$60.1\%$ are far below the $123.6\%$ tax (margins $-0.64$ to $-1.10$,
raw-float); on RTX4090 the same pricing rejects the serve top-1 (margin
$-8.568558$~pp) while still attaching at rank-3 ($+5.595340$~pp), so the tax
changes the decision, it does not blanket-prohibit.
On training the tax flips sign: $+26.6\%$ on RTX4090 becomes $-7.7\%$ on H800 ---
a negative tax, the generated kernel outrunning the Megatron python
SequentialMLP per-expert loop baseline on sm\_90.
The sign flip is not a contradiction: each phase is priced against its own
native baseline --- serving against the fast \texttt{fused\_moe}, training
against the slow python loop --- so identical generated kernels read as taxed in
one phase and subsidized in the other.
The negative tax carries a dual-baseline caveat (\cref{tab:tax} note): its
baseline is the python loop, but the searcher's cost model uses an optimized
grouped-GEMM loop curve --- the generated kernel beats the former, loses to the
latter (h800 fused/loop ${\approx}0.59$), so even a negative tax leaves the
searcher declining to attach ($0/5$). We anchor
the committed ep2 $-13.4\%$ / mean $-7.7\%$; a 3-round quick-verification (ep2
patched $62{,}894.4 >$ unpatched $57{,}336.8$, $+9.7\%$, same sign) corroborates,
with the unpatched baseline jittering ${\sim}{\pm}10\%$ across three shared-node
measurements (a single round once reversed the sign, which is why we require a
3-round median).

\subsection{RQ5: Ablation and discriminative power}

On the local grid the 24-cell ablation (2 phases $\times$ 3 refinement levels
$\times$ 2 constant sources $\times$ 2 max-violation points) is flat: within
every (phase, constants, max-vio) slice the L1/L2/L3 top-1 plan and predicted
throughput are bit-identical, with structural monotonicity ($\text{L2}\geq
\text{L1}$, $\text{L3}\geq\text{L2}$) holding by construction. This is a
property of the small local grid, not a searcher defect: training $pp{=}1$ makes
the schedule-refinement dimension structurally inert and the shape misses the
autotune grid (training \texttt{kernel=None} because data is missing,
\texttt{kernel\_no\_autotune\_data} $10$), while serving honestly declines the
kernel (\texttt{kernel=None} because the gain is below the $41.7\%$ tax, count
$0$).
Concretely the top-1 is constant at \texttt{tp1/pp1/ep1/1f1b+ov} for training
and \texttt{tp2/pp1/ep1/1f1b} for serving across all three refinement levels;
e.g.\ the calibrated max-vio $0.0$ slice holds train $20{,}085.3$ and serve
$456.5$ tok/s/GPU identically at every level.
The default-constants slices carry much higher predicted values (factory
optimism), so we draw no comparison from them; and even after P4b.6 expanded the
shape grid, local training still declines $0/5$ (staleness did not flip).

Discriminative power is demonstrated by swapping the workload. On
\texttt{kimi\_k2} (\texttt{k2\_train} phase) the corrected-accounting calibrated
slice moves $+8.11\%$ to $+8.51\%$ from L1 to L2 (max-vio $0.0$: $520.2 \to
562.4$ tok/s/GPU, $+8.11\%$; max-vio $0.2$: $453.1 \to 491.7$, $+8.51\%$) ---
nontrivial movement on the same harness, so the flatness is the grid, not the
searcher. \texttt{kimi\_k2} exercises the wide-EP, expert-heavy regime
(ep128 plans) that a local two-GPU grid structurally cannot reach, which is
precisely where the refinement levels start to separate.
\emph{Caliber:} these \texttt{kimi\_k2} numbers are a committed artifact with
generating-commit provenance (\texttt{ablation\_matrix\_kimi\_k2.json}, the
\texttt{k2\_train} oracle phase
under the rtx4090 overlay), on the same evidentiary footing as the rest of the
evaluation rather than a reproducible-by-call promise; the base
\texttt{ablation\_matrix.json} holds only the small-moe-bench and v2-lite grids.
The L1$\to$L2 gain is pure schedule refinement on a fixed
\texttt{tp1/pp2/ep128/interleaved+ov} layout.
The L1 plan itself changed under corrected accounting: the old L1
(\texttt{tp1/pp4/ep128/dualpipe+ov}) was ruled infeasible ($78.02 > 72$~GB) and
re-selected as \texttt{tp1/pp2/ep128/interleaved+ov}, so the flat grid still
reflects a live realizability decision.

\subsection{Honest negatives}

\cref{tab:negatives} indexes the honest negatives disclosed across this
evaluation --- from the terminal training FAIL and the unbuildable frontiers to
the four-fold launch-probe falsifications and the large-$me$ DeepEP losses.
Each entry is a first-class result with its own number, caliber, and
subsection; several (HN1, HN3, HN4) record negatives that a later, more capable
pipeline stage superseded and are preserved byte-for-byte as time-stamped
records rather than overwritten.
Read top to bottom, the negatives cluster on the two axes the paper claims are
hard --- realizability (HN1, HN14--HN16, HN19--HN20) and cross-generation kernel
economics (HN7, HN9--HN11, HN17) --- where an honest full-space optimizer should
be expected to fail, and does.

\begin{table}[t]
\centering
\caption{Success-criterion panorama (automatic $\geq \max(\text{hand-tuned})
\times 0.98$, raw-float). Throughput columns are aggregate tok/s \emph{except} the
$8{\times}$H800 train row, which is tok/s/GPU; ratios are unit-invariant. The four
verdicts are not same-source: both serving rows are argv-identical (H800's
$1.0304$ is same-configuration run-to-run noise, not a win over the hand-tuned
side); local training $+2.9\%$ is a margin against the $0.98{\times}$baseline
threshold ($+0.86\%$ vs.\ the baseline itself).}
\label{tab:panorama}
\footnotesize
\begin{tabular}{@{}llll@{}}
\toprule
\tblhead{Machine $\times$ phase} & \tblhead{Auto.} & \tblhead{Hand (max)} & \tblhead{Ratio / verdict} \\
\midrule
Local $2{\times}$4090 train  & $38{,}028.3$ & $37{,}703.4$ & $+2.9\%$ \textbf{PASS} \\
Local $2{\times}$4090 serve  & $3{,}510.9$  & $3{,}549.9$  & $0.9890$ \textbf{PASS} \\
$8{\times}$H800 serve         & $12{,}274.6$ & $11{,}912.5$ & $1.0304$ \textbf{PASS} \\
$8{\times}$H800 train         & $12{,}059.0$ & $12{,}913.2$ & $0.9338$ \textbf{FAIL} \\
\bottomrule
\end{tabular}
\end{table}

\begin{table}[t]
\centering
\caption{Preregistered adjudication, three branches over two phases $\times$ two
constant sources. m/p is measured/predicted; $\text{m/p}<1$ means the cost model
over-predicts throughput ($1/\text{m/p}$ is the over-prediction factor). The
opposite-direction local-serve magnitude gap (m/p $\approx 4.13$,
under-prediction) is HN8 in \cref{tab:negatives}.}
\label{tab:adjudication}
\footnotesize
\setlength{\tabcolsep}{3.5pt}
\begin{tabular}{@{}ll>{\raggedright\arraybackslash}p{0.40\columnwidth}@{}}
\toprule
\tblhead{Constants $\times$ phase} & \tblhead{Branch} & \tblhead{Key evidence} \\
\midrule
train $\times$ default    & honest-neg. & frontier $0/6$, nbhd.\ $0/4$; top-1 unrealizable \\
train $\times$ calibrated & honest-neg. & same; calibration changes numbers, not realizability \\
serve $\times$ default    & honest-neg. & top-1 unrealizable; m/p $0.285/0.301/0.323$ (strong \emph{and} weak validation both fail) \\
serve $\times$ calibrated & \textbf{STRONG} & top-1 match (\texttt{tp8}); m/p $0.738/0.749/0.848$; guardrailed \\
\bottomrule
\end{tabular}
\end{table}

\begin{table}[t]
\centering
\caption{Realization tax across generations (\texttt{overhead\_frac} $=
\text{unpatched}/\text{patched}-1$, applied as throughput $\div(1{+}\text{frac})$;
each leg is a 2-point arithmetic mean, no assumed sign). Negative-tax
dual-baseline note: the $-7.7\%$ is against the python SequentialMLP loop; the
searcher's own cost model uses an optimized grouped-GEMM loop curve
(fused/loop $\approx 0.59$), so the searcher still declines to attach the kernel
($0/5$, both machines).}
\label{tab:tax}
\footnotesize
\begin{tabular}{@{}llll@{}}
\toprule
\tblhead{Leg} & \tblhead{RTX4090 (sm\_89)} & \tblhead{H800 (sm\_90)} & \tblhead{Effect} \\
\midrule
serve & $+41.7\%$ & $+123.6\%$ & harsher; $0/5$ attach both \\
train & $+26.6\%$ & $-7.7\%$   & sign flip (negative tax) \\
\bottomrule
\end{tabular}
\end{table}

\begin{table*}[t]
\centering
\caption{Honest negatives index. Every negative is a first-class result
disclosed in-place in the cited subsection; time-stamped records are preserved
(supersede, not overwrite).}
\label{tab:negatives}
\footnotesize
\begin{tabular}{@{}llll@{}}
\toprule
\tblhead{\#} & \tblhead{Honest negative} & \tblhead{Key number} & \tblhead{Where} \\
\midrule
HN1  & train preregistered honest-negative (both const.\ sources) & \texttt{no\_measured\_train\_point} & RQ1 \\
HN2  & serve $\times$ default honest-negative; ranking wrong & m/p $0.285/0.301/0.323$ & RQ3 \\
HN3  & success crit.\ train v1: $0/7$ candidates run & pass$=$null & RQ2 \\
HN4  & success crit.\ train v2: $0$ probe-clean over 12 probes & first\_clean$=$None & RQ2 \\
HN5  & train terminal computable FAIL (one overlap flag) & $0.9338$ & RQ2 \\
HN6  & serve v1 FAIL (post-tax ranking tension; honest losing choice) & $0.8765$ & RQ2 \\
HN7  & P4a local realization tax (patched slower than native) & serve $41.7\%$ / train $26.6\%$ & RQ4 \\
HN8  & local serve magnitude gap widened under honest pricing & m/p $\approx 4.13$ & RQ3 \\
HN9  & DeepEP V2 2.1.0 unbuildable (three root causes) & CUDA${\geq}12.8$ needed & RQ4 \\
HN10 & large-$me$ regime loses to DeepEP & $-51.7\%$ at $me{=}1024$ & RQ4 \\
HN11 & cuda-graph structural conflict reproduced across generations & \texttt{sglang\_patch.py:177} & RQ4 \\
HN12 & 24-cell ablation flat (small-grid property) & 24 cells bit-identical & RQ5 \\
HN13 & kimi\_k2 old L1 ruled infeasible by corrected accounting & $78.02 > 72$~GB & RQ5 \\
HN14 & memory gate necessary-not-sufficient (ep4/ep8 over-admitted) & $57.8/47.0$ vs.\ ${\sim}74$~GiB & RQ1 \\
HN15 & FSDP memory prediction directly falsified & $69.32$ vs.\ ${\sim}78$--$79$~GiB & RQ1 \\
HN16 & offload/FSDP dual misjudgment in one task & --- & RQ1 \\
HN17 & negative tax but searcher still $0/5$ (dual-baseline mismatch) & fused/loop $\approx 0.59$ & RQ4 \\
HN18 & v3c $\alpha$-removal hypothesis falsified by v3d (no flip) & $+8.1\% \to +2.28\%$ & RQ2 \\
HN19 & frontier-scale models all memory-infeasible on $8{\times}$H800 & $640$~GB total HBM ($8{\times}80$) & \S6 \\
HN20 & \texttt{dualpipe\_startup\_div} has no calibration carrier & const.\ $8.0$ unfit & RQ1 \\
HN21 & local serve historically mis-ranked (correct answer earned) & $2{,}925.7 \to 3{,}510.9$ & RQ3 \\
HN22 & stream-overlap physically net-negative at modest shapes & $-2.09/-5.64$ hiding & \S6 \\
\bottomrule
\end{tabular}
\end{table*}

%% file: sections/06-discussion.tex
\section{Discussion}\label{sec:discussion}

\begin{figure}[t]
\centering
\input{figures/fig4-checklist}
\caption{A reusable preregistration discipline for prediction-centric systems
evaluation. Predictions are frozen and the generating artifact hashed before
any measurement; adjudication sorts each prediction into a strong match, a weak
frontier agreement, or an honest negative; every revision is an append-only
amendment that never rewrites the frozen prediction; the amended model is
re-preregistered before the next round.}
\label{fig:checklist}
\end{figure}
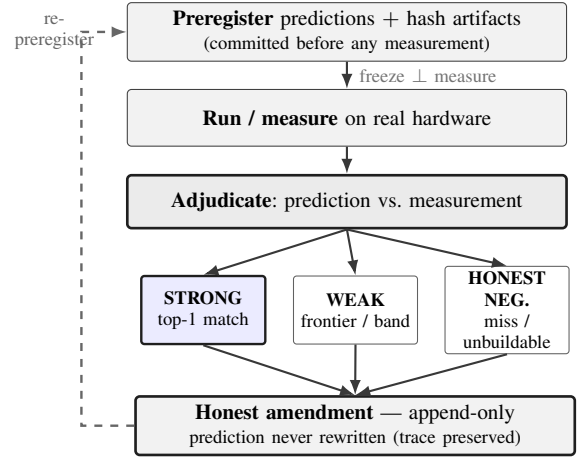

\subsection{A reusable preregistration discipline}

Our evaluation rests on a discipline that is independent of \sys and, we
believe, transferable to any systems study whose central claim is a prediction
(\cref{fig:checklist}). The loop has five steps. \emph{(i)~Preregister}: commit
the predicted ranking and hash the generating artifact before any measurement
touches the hardware. \emph{(ii)~Measure} on the real machine.
\emph{(iii)~Adjudicate} the prediction into one of three branches --- a strong
match (predicted top-1 equals measured top-1), a weak frontier/band agreement, or
an honest negative (a miss, or an unbuildable plan). \emph{(iv)~Amend}
append-only: a corrected cost model is an amendment that never rewrites the frozen
prediction, so the trace from first guess to final verdict survives
byte-for-byte. \emph{(v)~Re-preregister} before the next round. We ran this across
ten amendments on 8$\times$H800; every revision was appended and round-trip
verified, none edited after the fact.
The procedure costs little to adopt and makes an optimizer's honest negatives
auditable rather than deniable.

\subsection{How the boundaries generalize}

The eight toolchain boundary layers of \cref{tab:boundaries} are not
idiosyncratic to our cluster; they fall into a few classes that recur across
systems and grow \emph{harsher} with scale. Layer-count divisibility (an
interleaved schedule requires $pp\cdot v \mid 27$), sharding-semantics accounting
(expert-DP optimizer sharding, FSDP working buffers), and data-dependent dynamic
shapes (the cuda-graph capture conflict) are all properties of the toolchain,
not of eight GPUs.
A larger world size multiplies the divisibility constraints that must hold
simultaneously and enlarges the sharding-accounting surface, so the problem the
predicate solves gets bigger, not smaller. The mechanism is scale-invariant by
construction --- the predicate reuses the production emitter as an \texttt{emit}
dry-run oracle, inheriting exactly the constraints the real toolchain enforces
--- and predicate plus search \emph{run} over the published 2048-GPU
configuration (256~nodes~$\times$~8~H800) in $70$~ms (\cref{sec:eval}). We are
precise: this shows the predicate and search execute on that configuration at low
cost, not that we measured end-to-end throughput at 2048 GPUs. Whether predicate
\emph{coverage} degrades under multi-node-specific failure modes (topology-aware
collectives, cross-node fabric) is untested and registered open; the claim is
mechanistic and cost-based, not an empirical scaling result.
One might still object that realizability is a sideshow and accuracy is the true
bottleneck. Our own data refutes the reversal: domain-scoped calibration
repaired the serving ranking to a strong, hardware-reproduced match, yet the
predicted training frontier stayed 0 of 6 buildable --- fixing the numbers did
not make the plans buildable.
And the phenomenon only surfaces once the loop is closed: a search that stops at
a ranked plan, or a codegen backend handed a fixed strategy, never observes that
its top-ranked plan is unbuildable. This is why we treat the two spaces as
distinct objects rather than as one noisy estimate of the other. The gap between the space the cost model
ranks and the space the toolchain can build is not merely unclaimed ground --- it
is visible only from inside a closed loop.

\subsection{Open problems}

Our machine-readable debt list carries 31 open entries;
three are representative, and each sits at a different tier of the search. First,
a \emph{CUDA-C++ codegen backend}: a
device-initiated NVSHMEM spike proved the toolchain path (SMOKE\_PASS), but stock
Triton~3.3.0 exposes no relocatable-device-code or device-linking route, so
emitting cross-rank communication kernels needs a CUDA-C++ escape hatch.
Second, \emph{communication-pattern asymmetry}: the all-to-all versus all-gather
overlap left a $+5.05\%$ residual credit that our schedule model does not yet
attribute.
Third, the \emph{Hopper large-$m_e$ regime}: past the crossover a hand-tuned
cross-rank library outruns our generated kernels by up to $51.7\%$
(\cref{fig:crossover}); mapping where generated kernels should defer to such
libraries is open.
None of the three changes the realizability contract: each would widen the
buildable space the predicate can admit, which is precisely the direction the
design is meant to grow.

\subsection{Limitations and threats to validity}

We state the threats ourselves, and for each we bound what it does and does not
undermine. \emph{Single node, small models.} We evaluated on one 8-GPU node and
models that fit it; multi-node topology search is registered as open.
This bounds the external validity of the throughput numbers, not the
realizability thesis --- the predicate's verdicts are computed from the same
emitter a production run uses, at any scale. \emph{Parity, not superiority.} On
both serving generations the automatic and hand-tuned commands are
argv-identical; the 8$\times$H800 serving ratio of 1.0304 is same-configuration
run-to-run noise, not the automatic side out-running the manual side.
The parity is nonetheless earned: the default constants mis-ranked every serving
plan (over-prediction $3.1$--$3.5\times$) and only domain-scoped calibration
recovered the hardware-correct order.
\emph{Large-$m_e$ kernels lose to DeepEP.} Our generated expert kernels win
per-token below the crossover but lose above it, by $51.7\%$ at $m_e{=}1024$;
we make no claim to beat a hand-tuned communication library, and report the
crossover as the finding. \emph{Training fails its success criterion.} On
8$\times$H800 the automatic training plan reaches only 0.9338 of the best
hand-tuned throughput, a terminal honest FAIL below the 0.98 bar.
The failure is one scheduling flag on an otherwise identical offloaded plan and
is a \emph{computable} negative --- the plan builds and runs --- categorically
apart from the unbuildable-plan negatives that motivate the paper.
\emph{Shared-node noise.} The H800 node was shared; round-to-round spread reached
$\sim$28\% on one serving plan and two external co-tenant events were logged.
We report three-round medians and apply the 0.98 rule on raw floats; the verdicts
follow the preregistered rule rather than a claim of statistical robustness,
and on the training FAIL the closest estimator's margin lies within the disclosed
noise band, which we say plainly (\cref{sec:eval}).
\emph{Launch probe is a campaign practice, not a shipped loop.} The mandatory
launch probe of P3 is here a manual/semi-automatic walk-down we ran during the
campaign (\texttt{automatic\_walk\_provenance}), not a component the shipped
\texttt{search()} invokes: a new user obtains a predicate-passed ranked list and
must still walk-and-launch for the launch guarantee. A callable
\texttt{walk\_and\_launch} loop is open.
\emph{Calibration is within-hardware and within-model.} The serving strong match
fits constants from H800 microbenchmarks and adjudicates them against H800
end-to-end serving of the \emph{same} model --- distinct measurements, but shared
hardware and model, with no cross-workload holdout on the serving side (the
RTX4090 training calibration does hold one configuration out, at $2.6\%$).
\emph{One model family.} Aside from a single \texttt{kimi\_k2} ablation slice,
every measured number is a v2-lite-class reference architecture; the
frontier-scale models that motivate the work are memory-infeasible on one
8$\times$H800 node (HN19), so the realizability thesis is argued at their scale
but the throughput numbers are not.
\emph{Toolchain version drift.} The necessary-condition guarantee is relative to
the pinned Megatron \texttt{core\_v0.18.0} and SGLang \texttt{0.5.15}: an upgrade
can add or remove guards, so a verdict is versioned --- the predicate re-derives
it for free by re-running the emitter, but old verdicts are not auto-revalidated.
A full index of these and the remaining honest negatives is \cref{tab:negatives}.

%% file: figures/fig4-checklist.tex
\begin{tikzpicture}[
  font=\footnotesize,
  node distance=3.6mm,
  pbox/.style={rectangle, rounded corners=1.6pt, draw=black!75, fill=black!5,
               align=center, inner sep=3pt, text width=56mm, minimum height=7.5mm},
  adj/.style={rectangle, rounded corners=1.6pt, draw=black!88, line width=1.0pt,
              fill=black!8, align=center, inner sep=3pt, text width=56mm,
              minimum height=7mm},
  br/.style={rectangle, rounded corners=1.2pt, draw=black!62, fill=white,
             align=center, inner sep=2pt, text width=15mm, minimum height=9mm,
             font=\scriptsize},
  brpos/.style={br, fill=blue!8, draw=black!88, line width=0.9pt},
  amend/.style={rectangle, rounded corners=1.6pt, draw=black!88, line width=1.0pt,
                fill=black!5, align=center, inner sep=3pt, text width=56mm,
                minimum height=7.5mm},
  flow/.style={-{Latex[length=2mm]}, draw=black!78, line width=0.8pt},
  loop/.style={-{Latex[length=2mm]}, draw=black!60, line width=0.9pt, dashed},
]

\node[pbox] (pre)
  {\textbf{Preregister} predictions $+$ hash artifacts\\{\scriptsize(committed before any measurement)}};
\node[pbox, below=of pre] (run)
  {\textbf{Run / measure} on real hardware};
\node[adj, below=of run] (adjn)
  {\textbf{Adjudicate}: prediction vs.\ measurement};

\node[brpos, below=6mm of adjn.south, xshift=-19mm] (strong)
  {\textbf{STRONG}\\top-1 match};
\node[br, right=3.5mm of strong] (weak)
  {\textbf{WEAK}\\frontier / band};
\node[br, right=3.5mm of weak] (neg)
  {\textbf{HONEST}\\\textbf{NEG.}\\miss / unbuildable};

\node[amend, below=6mm of $(strong.south)!0.5!(neg.south)$] (amend)
  {\textbf{Honest amendment} --- append-only\\{\scriptsize prediction never rewritten (trace preserved)}};

\draw[flow] (pre) -- node[right=1pt, font=\scriptsize, black!55]
        {freeze $\perp$ measure} (run);
\draw[flow] (run) -- (adjn);
\draw[flow] (adjn.south) -- (strong.north);
\draw[flow] (adjn.south) -- (weak.north);
\draw[flow] (adjn.south) -- (neg.north);
\draw[flow] (strong.south) -- (amend.north);
\draw[flow] (weak.south) -- (amend.north);
\draw[flow] (neg.south) -- (amend.north);

\draw[loop] (amend.west) -| ($(pre.west)+(-6mm,0)$)
        -- node[left, font=\scriptsize, black!60, align=center]
        {re-\\preregister} (pre.west);

\end{tikzpicture}

%% file: sections/07-related.tex
\section{Related Work}\label{sec:related}

The literature this work touches divides along two seams. On the training
side, automatic-parallel frameworks search over parallel configurations; on
the serving side, large-scale expert-parallel deployments are assembled and
tuned by hand. The two stacks rarely share machinery, and neither treats
whether a chosen plan is buildable as an explicit search concern.
\Cref{tab:matrix} places \sys{} against fifteen representative systems along
five axes: parallel-strategy search, communication scheduling and overlap,
kernel code generation, unified training and serving, and deployment
realizability. The first four axes are established design concerns; the
fifth---whether a top-ranked plan can actually be built by the target
toolchain---is the concern this paper isolates. No prior system spans all
five, and the three axes that \sys{} closes jointly (search, kernel
generation, and train/serve unification) are the ones that no single line of
work has connected end to end.

\begin{table*}[t]
\centering
\caption{Capability matrix: \sys{} against representative MoE and
automatic-parallel systems along five axes.
\yes{}~full, partial, \no{}~none or unaddressed, --~not applicable.}
\label{tab:matrix}
\small
\begin{tabular}{@{}lccccc@{}}
\toprule
System & \tblhead{Strategy search} & \tblhead{Comm.\ overlap} & \tblhead{Kernel codegen} & \tblhead{Train+serve} & \tblhead{Realizability$^\dagger$} \\
\midrule
Alpa~\cite{zheng2022alpa}                       & \yes    & \no      & partial & \no     & \no \\
SmartMoE~\cite{zhai2023smartmoe}                & \yes    & \no      & \no     & \no     & \no \\
Galvatron~\cite{galvatron2023}                  & \yes    & partial  & \no     & \no     & \no \\
Metis~\cite{um2024metis}                        & \yes    & \no      & \no     & \no     & \no \\
nnScaler~\cite{lin2024nnscaler}                 & \yes    & partial  & partial & \no     & \no \\
Lancet~\cite{lancet2024}                        & \no     & \yes     & \no     & \no     & \no \\
Tutel/Flex~\cite{tutel2023}                     & partial & partial  & \no     & \yes    & \no \\
FlexMoE~\cite{flexmoe2023}                      & partial & \no      & \no     & \no     & \no \\
TileLink~\cite{zheng2025tilelink}               & \no     & \yes     & \yes    & --      & \no \\
Triton-distributed~\cite{zheng2025tritondist}   & \no     & \yes     & \yes    & partial & \no \\
FlashMoE~\cite{flashdmoe2025}                    & \no     & \yes     & \no     & \no     & \no \\
SGLang large-EP~\cite{sglang2025largeep}        & \no     & partial  & \no     & \no     & \no \\
TRT-LLM Wide-EP~\cite{nvidia2026wideep}         & \no     & partial  & \no     & \no     & \no \\
UniEP~\cite{zheng2026uniep}                      & \no     & \yes     & \yes    & \no     & \no \\
DITRON~\cite{ditron2026}                        & \no     & \yes     & \yes    & \yes    & \no \\
\textbf{\sys{} (this work)}                      & \yes    & \yes     & \yes    & \yes    & \yes \\
\bottomrule
\end{tabular}

\vspace{2pt}
{\footnotesize $^\dagger$~Realizability is a dimension this paper introduces; marks unaddressed concerns, not a scoring of prior work.}
\end{table*}

\paragraph{Three nearest neighbors}
Each of \sys{}'s three search axes has a close single-axis precedent, but none
reaches the opposite end of the loop. \textbf{SmartMoE}~\cite{zhai2023smartmoe}
is the only system that treats expert parallelism and expert placement as
first-class search dimensions, pairing an offline data-sensitive analytic cost
model with a lightweight online plan selector; it selects among existing
kernels and collectives, with no kernel generation, no overlap compilation,
and no serving path, so its coverage is the strategy-search axis alone.
\textbf{nnScaler}~\cite{lin2024nnscaler} is the only surveyed auto-parallel
framework that emits executable code, converting partitioned subgraphs back
into a runnable PyTorch program; but its search carries no MoE or
expert-parallel dimension, the emitted artifact stops at graph-level PyTorch
(kernel fusion is declared complementary and out of scope), and it targets
training only. \textbf{Lancet}~\cite{lancet2024} is the only MoE-specific
whole-graph computation--communication overlap compiler; yet its parallel
strategy is a fixed input (automatic sharding is left as future work), it
reuses existing kernels rather than generating them, and its backward-pass
overlap has no serving analogue. The three neighbors thus partition \sys{}'s
axes one apiece, and none simultaneously touches both strategy search and
GPU-kernel generation. In each case the optimizer ranks plans it cannot
itself emit as running kernels, so the buildability of the top choice is
never part of the objective.

\paragraph{Closest concurrent work}
Two recent systems from a single group come nearest.
\textbf{UniEP}~\cite{zheng2026uniep} extends a distributed compiler to fuse
dispatch, grouped-GEMM, and combine into a training MegaKernel and autotunes
over a large space of kernel configurations; it is a genuine precedent for
automatically generating and tuning MoE expert-parallel kernels, which we do
not claim to originate. Its expert-parallel degree and parallel combination
are, however, fixed inputs---the paper states it is orthogonal to placement
strategies---and it targets training only. \textbf{DITRON}~\cite{ditron2026},
from the same team, is a distributed tile-level compiler that generates
compute and communication kernels for both training and inference, but is not
MoE-specific and performs no strategy search. Read together, these two lines
cover strategy-independent kernel generation plus train/serve breadth; what
remains open is joining them to a search over parallel strategies, and doing
so under a realizability constraint---the two axes on which \sys{} does not
overlap them. We accordingly position UniEP-style kernel tuning as a reusable
backend below \sys{}'s code generator rather than as a competitor. Because
these lines share authorship with the nearest strategy-search precedents, we
regard this team as the most likely source of future gap-narrowing and state
the boundary plainly rather than working around it.

\paragraph{Neighborhood}
The kernel and code-generation ecosystem supplies backends rather than
optimizers. \textbf{TileLink}~\cite{zheng2025tilelink} provides tile-centric
primitives and a compilation framework for a single layer's overlap under a
fixed parallelization strategy, and does not search over strategies;
\textbf{Triton-distributed}~\cite{zheng2025tritondist} compiles overlapping
kernels with a tiling autotuner but leaves the parallel strategy
predetermined; \textbf{FlashMoE}~\cite{flashdmoe2025} hand-writes a single
persistent inference kernel; and
\textbf{MegaBlocks}~\cite{megablocks2022} and
\textbf{DeepGEMM}~\cite{deepseek2025deepgemm} supply block-sparse and
grouped-GEMM kernels. The communication line---%
\textbf{Comet}~\cite{comet2025}, \textbf{MegaScale-MoE}~\cite{megascalemoe2026},
and \textbf{DeepEP}~\cite{deepseek2025deepep}---optimizes MoE overlap and
expert-parallel transport under a given strategy. Automatic-parallel planners
\textbf{Alpa}~\cite{zheng2022alpa}, \textbf{Metis}~\cite{um2024metis}, and
\textbf{Galvatron}~\cite{galvatron2023} search dense parallel spaces without a
first-class EP/MoE dimension or kernel-level code generation. Cost-model and
benchmarking neighbors \textbf{Vidur}~\cite{vidur2024} and
\textbf{MoE-CAP}~\cite{moecap2024} rank or measure configurations but do not
build them, sharing the very ranking-versus-building gap that our realizability
axis targets. Concurrent rack-scale and MoE-strategy efforts---%
\textbf{UltraEP}~\cite{ultraep2026}, \textbf{UCCL-EP}~\cite{ucclep2026}, and
\textbf{TeleChat3-MoE}~\cite{telechat2025}---each still miss at least one axis
(kernel generation and search; code generation and search; and code generation
and serving, respectively).

\paragraph{Methodology}
We pre-register ranked predictions in a frozen adjudication file before
executing on the H800 testbed and report the outcomes---including where our own
predictions are falsified---as-is (\cref{sec:eval,sec:discussion}); we make no
claim about the rarity of this practice.

Across all of these systems, deployment realizability is uniformly
unaddressed, and this is not incidental. The gap \sys{} occupies is
conjunctive: only when parallel-strategy search, communication scheduling,
kernel code generation, and unified training and serving are closed into one
loop does the mismatch between the space a cost model ranks and the space a
toolchain can build become observable at all. That vantage point, not any
single axis, is what \sys{} adds.

%% file: sections/08-conclusion.tex
\section{Conclusion}\label{sec:conclusion}

A cost model ranks plans in one space; a toolchain can build plans in another,
and the two are not the same. \sys treats that gap as a first-class search
constraint: a realizability predicate reuses the emitter as an \texttt{emit}
dry-run oracle, a memory gate and a mandatory launch probe (a walk-down in our
evaluation, in-searcher automation left as future work) catch what static
reasoning misses, and a realization tax prices realizable-but-costly plans
rather than forbidding them. Across two hardware generations and both training
and serving, the search reaches the hand-tuned frontier where a plan is
buildable --- a local training win of $+2.9\%$ above the $0.98\times$
acceptance bar
and serving parity on 8$\times$H800 (ratio $1.0304$)
--- and it characterizes, rather than hides, the places where it is not. We
report every honest negative under a preregistered, artifact-hashed adjudication
whose predictions are frozen before measurement and revised only by append-only
amendment, including the 6.6 points we lose on 8$\times$H800 training (a terminal
ratio of $0.9338$).
We argue that for an MoE full-space optimizer, buildability belongs in the
objective, not in a post-hoc filter: realizability is a first-class citizen of
the search.

%% file: sections/09-appendix.tex
\section{Measurement Calibers and Additional Detail}\label{sec:appendix}

This appendix expands the caliber boxes referenced from \cref{sec:eval}.

\paragraph{Testbeds and calibration domains}
The local node is $2{\times}$RTX4090 (sm\_89); the remote node is
$8{\times}$H800 (sm\_90) with an all-to-all NVLink (NV8) topology. Calibrated
constants are domain-scoped: the H800 domain is
$\{\text{gpu\_name}{:}\ \text{NVIDIA H800},\ \text{dtypes}{:}\ [\text{bf16}],\
m\_\text{range}{:}\ [64, 8192]\}$, and the RTX4090 domain shares the same
$m$-range; an unset (default) constant leaves the prior behavior bit-identical.
The H800 kernel campaign ran on \texttt{torch} 2.7.0+cu126 and
\texttt{triton} 3.3.0.

\paragraph{Success rule and thresholds}
The success criterion is automatic $\geq \max(\text{hand-tuned}) \times 0.98$
applied to raw floats. The terminal H800 training FAIL is reported at three
estimators, all below $0.98$: $0.9338$ (formal 3-round median), $0.9690$
(confirmation round), and $0.9517$ (four-round merged).
The local \texttt{moefs gate} passes at train $38{,}028.3 \geq 36{,}949.3$
($=0.98 \times 37{,}703.4$) and serve $3{,}510.9 \geq 3{,}478.9$
($=0.98 \times 3{,}549.9$), exit $0$.
H800 artifacts do \emph{not} enter the \texttt{moefs gate} judgment, which is
anchored to the local $2{\times}$RTX4090; the local serve threshold
($3{,}478.9$) is supplied by the amendment text, since
\texttt{serve\_results.json} itself carries no verdict/threshold field (the gate
recomputes it at run time).

\paragraph{Protocols}
The kernel crossover (\cref{fig:crossover}) uses CUDA-event timing, warmup $10$,
$200$ reps for $me<16$ and $50$ otherwise, and 3-round median-of-medians, at
shape $E{=}64,K{=}2048,N{=}1408,\text{top\_k}{=}2$ with \texttt{world\_size} $8$
and $8$ local experts; the DeepEP baseline is v1.2.1 (NVSHMEM).
The realization tax (\cref{tab:tax}) measures serving over 3 rounds
$\times$ 120\,s at concurrency $32$ and training over 3 rounds $\times$ 30 steps
(median), and the overlap penalty is anchored at $0.027086$.
Two Python environments (3.12 and 3.13) were used and their differences are
disclosed; the SGLang fused single-GEMM baseline runs in an isolated venv.

\paragraph{Rounding and provenance caveats (U8--U11)}
Overlap hiding-ratios are the precise $0.3147$ and $0.5073$ (README rounds
$0.315/0.507$); an earlier P3b prose value ($0.41$--$0.55$ at chunks $4$)
differs by measurement jitter (spread $0.061$) and is superseded by the current
values.
From amendment~4 onward, \texttt{appended\_at\_commit} records the
generating-tree provenance, with the byte-level landing one commit later (an
explicit erratum note; amendments 5--10 follow the same convention).
The v2-lite training \texttt{fused\_by\_shape} grid samples
$me \in \{256, 512, 1024, 2048\}$; the value $0.2316$ cited at $me{=}768$ is the
committed fused curve (\texttt{eff\_max} $0.2856$) evaluated at $me{=}768$, not
a raw sample.
Shared-node co-tenancy is disclosed: the \texttt{tp4} inter-round range reaches
${\sim}28\%$, and GPU3 saw two external tenants ($32.66$ and $16.65$ GiB) during
the campaign; the idle gate covers launch only, and serving was collected across
sessions.

\paragraph{Artifact availability}
The code and every artifact this paper cites---the \texttt{moefs} package, the
\texttt{artifacts/**.json} files, the frozen \texttt{h800\_adjudication.json}
with its \texttt{sha256}-locked append-only amendments, and the two artifacts
produced for this revision (\texttt{search\_cost.json},
\texttt{ablation\_matrix\_kimi\_k2.json})---will be released with the paper, so
that each \texttt{\%~SRC} pointer resolves to an openable file and the phase
gate recomputes for any reader. The release URL and license are pending author
confirmation: \texttt{<repository URL pending>}.